\documentclass[10pt]{article}

\usepackage{graphicx}
\usepackage{color}
\usepackage{natbib}
\usepackage{amsmath}

\newcommand{\kmps}{\;{\rm km\;s^{-1}}}
\newcommand{\cmmt}{\;{\rm cm^{-3}}}
\newcommand{\gcmmt}{\;{\rm g\;cm^{-3}}}

\begin{document}

\title
{Modelling stellar jets with magnetospheres\\
  using as initial states analytical MHD solutions}

\maketitle

\author
{
{\bf
 P.~Todorov$^1$, T.~Matsakos$^{2,3}$, V.~Cayatte$^1$, C.~Sauty$^1$, 
 J.J.G.~Lima$^{4,5}$, K.~Tsinganos$^{6,7}$ }
\\
\bigskip

\noindent
$^1$ LUTh, Observatoire de Paris, UMR 8102 du CNRS, Universit\'e Paris Diderot, F-92190 Meudon, France \\
$^2$ Department of Astronomy \& Astrophysics, The University of Chicago, Chicago, IL 60637, USA \\
$^3$ LERMA, Observatoire de Paris, Universit\'e Pierre et Marie Curie, Ecole Normale Sup\'erieure, Universit\'e Cergy-Pontoise, CNRS, France \\
$^4$ Instituto de Astrof\'{\i}sica e Ci\^{e}ncias do Espa\c{c}o, Universidade do Porto, CAUP, Rua das Estrelas, 4150-762 Porto, Portugal \\
$^5$ Departamento de F\'{\i}sica e Astronomia, Faculdade de Ci\^{e}ncias, Universidade do Porto,
Rua do Campo Alegre 687, 4169-007 Porto, Portugal \\
$^6$ Department of Astrophysics, Astronomy and Mechanics, Faculty of Physics, University of Athens, 15784 Zografos, Athens, Greece \\
$^7$ National Observatory of Athens, Lofos Nymphon, Thission 11810, Athens, Greece}

\begin{abstract}
In this paper we focus on the construction of stellar outflow models emerging from a polar coronal hole-type region surrounded by a 
magnetosphere in the equatorial regions during phases of quiescent accretion.  The models are based on initial analytical solutions.
We adopt a meridionally self-similar solution of the time-independent and axisymmetric MHD equations which describes effectively a jet originating from 
the corona of a star. We modify appropriately this solution in order to incorporate a physically consistent stellar magnetosphere.
We find that the closed fieldline region may exhibit different behaviour depending on the associated boundary conditions and the distribution of the heat 
flux. However, the stellar jet in all final equilibrium states is very similar to the analytical one prescribed in the initial conditions.
When the initial net heat flux is maintained, the magnetosphere takes the form of a dynamical helmet streamer with a quasi steady state slow 
magnetospheric wind. With no heat flux, a small dead zone forms around the equator, surrounded by a weak streamer, source of a low mass flux magnetospheric 
ejection. An equilibrium with no streamer and a helmet shaped dead zone is found when the coronal hole is larger and extends up to a reduced 
closed fieldline static zone.
The study demonstrates the robustness of self-similar solutions and their ability to provide quasi-steady jet simulations with realistic boundaries and 
surrounded by equatorial magnetospheres. 
\end{abstract}

\maketitle

\section{Introduction}

Jets from low mass protostars have been extensively observed (e.g. \citeauthor{Cabrit07}, \citeyear{Cabrit07}; \citeauthor{Ferreira13}, \citeyear{Ferreira13};
\citeauthor{Franketal14}, \citeyear{Franketal14}). The accretion-ejection mechanism is expected to occur during almost all the pre-main sequence
stellar (PMS) evolution, but it is most extensively observed in jets associated with class II T Tauri stars. For these systems there is a wealth of astronomical data available
at different length scales and wavelengths, for both, the disc and the stellar jet of T Tauri stars (see \citeauthor{Franketal14}, \citeyear{Franketal14}, and references 
therein).

However, accretion may not be continuous while on dynamical time scale the stellar jet may remain steady. In other words, the jet may be continuous even during periods where accretion onto the star is episodic.
RY Tau is a 
good example of a faint jet with a low mass loss rate which may not be associated with an accretion episode (\citeauthor{Sautyetal11}, \citeyear{Sautyetal11}). Propellers 
are also a good example where there is no accretion onto the star but nevertheless there is a stellar wind. Of course in this case, the accreted mass is expelled in the X-
wind, see \cite{Lovelaceetal14}.

T Tauri jets are interesting and challenging also from the theoretical point of view because they exhibit a complex circumstellar system.
A disc wind emerges from the accretion disc helping the accretion process by removing angular momentum.
Accretion onto the star is believed to take place via accretion columns along a magnetosphere, i.e. the zone where the stellar magnetic fieldlines are
closed. The inner part of the magnetospheric structure is probably static, a zone resembling to the helmet streamers of our Sun.
The magnetospheric connection with the disc is thought to trigger reconnection processes and drive an X-type or magnetospheric wind, fed with mass 
mainly from the accretion disc.
Through this mechanism it is likely that part of the magnetic field of the star is  also carried away forming a streamer, a 
structure comparable to what is observed in our Sun.

Moreover the multi-component jet scenario 
 is both observationally and theoretically supported in T Tauri stars. The presence of a radial outflow
coming from the polar regions of the star coupled with a wind launched from the disc at a constant angle, is confirmed by the studies of He I $\lambda 10830$ profiles 
(\citeauthor{Edwards06}, \citeyear{Edwards06}; \citeauthor{Kwanetal07},  \citeyear{Kwanetal07}). 
However, long term observations are necessary to determine that there is a stellar jet related to the accretion shock. 
For  TW Hya, \cite{Dupreeetal2014}  inferred a high temperature for the wind, a result of the deposition of energy due to accretion. 
The individual contribution of the stellar axial outflow component and the disk wind component to the total outflow, is expected to depend 
on the intrinsic physical properties of the CTTs (Petrov et al., \citeyear{Petrovetal14}).

On the theoretical front, the various aspects of this complex problem have been studied separately, starting from analytical models which have then been extended via 
numerical simulations.
The original \cite{BlandfordPayne82} radially self-similar model for disc winds has been extensively studied and extended to 
include accretion. For a review of disc wind solutions see \cite{Stuteetal14} and references therein.
Several simulations have been performed for disc winds by various groups supporting the magneto-centrifugal mechanism for 
launching the wind from the disc (e.g. Tzeferacos et al. \citeyear{Tzeferacosetal09, Tzeferacosetal13}). 
However, to fit the observational data and specifically the observed jet radii (\citeauthor{Ferreiraetal06},  
\citeyear{Ferreiraetal06}), a warm corona on the disc is necessary.
Matsakos et al. (\citeyear{Matsakosetal08, Matsakosetal09, Matsakosetal12}) showed that analytical disc wind solutions are numerically stable, despite the modifications  
required to extend the self similar models to the axis where they are divergent.

Another set of analytical and numerical modelling has been put forward to study the connection of the magnetosphere with the disc.
The original so-called X-wind model by Shu and collaborators (e.g. Shu et al. \citeyear{Shuetal94}; Shang et al. \citeyear{Shangetal02}), 
has been widely exploited, especially numerically.
Several simulations (e.g. Fendt, \citeyear{Fendt09}; Romanova et al., \citeyear{Romanovaetal09a, Romanovaetal09b}; Zanni \& Ferreira,
\citeyear{ZanniFerreira09, ZanniFerreira13}; Ferreira \& Casse, \citeyear{FerreiraCasse13}) address the question of the interaction of the disc
with the dipolar field of the magnetosphere. In general, those simulations include the accretion disc and the disc wind. 
They have shown that, apart from the expected accretion columns, part of the disc mass can be expelled 
following magnetospheric reconnection. This is only achieved in a strongly time dependent way. 
Such a mechanism may explain flaring and violent ejections but cannot 
account for the overall continuous flow observed in many objects. 

The central axial stellar jet has also been studied analytically and numerically. Meridional self-similar models (see Sauty \& Tsinganos \citeyear{ST94, STT04}) provide a 
physical insight on the pressure driven mechanism for the launching of the central jet. This kind of models can be applied to 
T Tauri micro jets with low mass loss rates as in RY Tau (see also Zanni \& Ferreira, \citeyear{ZanniFerreira13}). The energetic problem usually invoked for stellar 
thermally driven winds in CTTS (e.g. Cabrit, \citeyear{Cabrit07}) may not hold as high thermal X ray 
temperatures have been detected by the Chandra Observatory (G{\"u}del et al., \citeyear{Gudeletal11}; Skinner et al., \citeyear{Skinneretal12}). This thermal X-ray 
emissions imply a strong coronal heating comparable to the one of the fast solar wind. Of course such stellar jet solutions cannot account for jets with 
high mass loss rates but they are an essential component of the overall outflow, inside the disc wind.  

It is worth to recall that stellar jets are the corner stone for solving the angular momentum problem and explaining the low rotation of T Tauri stars 
(Matt \& Pudritz, \citeyear{MattPudritz08a, MattPudritz08b}; Sauty et al., \citeyear{Sautyetal11}). It turns out that the disc-locking mechanism is rather inefficient 
in removing angular momentum (Zanni \& Ferreira, \citeyear{ZanniFerreira09}) while the stellar jet can very efficiently brake the star. 

Meridionally symmetric self-similar MHD solutions provide an excellent framework to study stellar jets and the solar wind.
Although they have been shown to be structurally stable at large distances (Matsakos et al., \citeyear{Matsakosetal08}), the configuration of the inner closed 
fieldline region is not physically consistent (Sauty et al., \citeyear{Sautyetal11}).
In particular, despite the fact that the assumption of self-similarity is appropriate to describe the outflow geometry and dynamics,
it also includes a flow inside the equatorial closed magnetic field lines.
Since the closed fieldlines thread the equatorial plane, it would imply an unphysical accumulation of mass there.
For the solar wind, \cite{TsinganosLow89}
proposed a sophisticated way to treat the dead zone in a self-consistent way.
However, adapting this method to the self-similar solution of \cite{Sautyetal11} is a rather difficult task because the geometry of the 
magnetosphere is not fixed a priori but is instead deduced from the solution itself.

The main goal of this paper is to improve our previous models insofar the behavior of the solution around the equator is concerned, 
by incorporating to a polar stellar jet, an equatorial magnetosphere explicitly in the picture. 
Specifically, we prescribe realistic {\it initial} and {\it boundary conditions} in the magnetosphere, while keeping the
analytical self-similar solution for the stellar jet emerging from the polar coronal hole. 
However, after this modification the initial magnetosphere is out of equilibrium with the rest of the solution. 
Hence, we perform numerical simulations to allow the system to relax to a new steady-state.
We study three different cases by assuming different initial and boundary conditions for the magnetosphere, focusing also on the 
distribution of the heat flux.
We explore the three possibilities of the formation of either just a dead zone, or only streamers, or a combination of the two.

This paper continues our efforts to self-consistently study the formation
of a polar jet with open fieldlines surrounded by an equatorial magnetosphere.
In the present effort, accretion is not taken into account.
Nevertheless, this work serves as a first step that aims to focus on
understanding the dynamics and interaction of the polar stellar jet and the
equatorial magnetosphere, disentangling them from any effects that accretion
might have.
In a subsequent work, we plan to incorporate accretion columns in order
to clarify their role in the evolution of the stellar jet.

\section{Polar outflows surrounded by streamers and dead zones}

\subsection{The initial setup of the analytical outflow solution}

We aim at finding global quasi-steady MHD structures of multicomponent jets, starting and building on the \cite{Sautyetal11} solution.

The stellar wind component of the initial analytical solution we use can model by itself the whole stellar jet of RY Tau. 
To summarize, the mass loss rate, polar asymptotic speed, 
stellar mass, radius, polar rotation frequency and equatorial angular speed are given, respectively, by
\begin{itemize}
\item $\dot M_{\rm wind} = 3.1\times 10^{-9}M_\odot/$yr\,,
\item $V_{z, \infty}= 393\kmps$\,.
\item ${\cal M} = 1.5\,M_\odot$\,,
\item $r_o = 2.4\,r_\odot$\,,
\item $\Omega = 5.99\,\, 10^{-6}$ rad/s or $V_{\varphi,o}= 10\kmps$\,.
\end{itemize}
These values are similar to the ones of RY Tau.
For this solution,  the physical quantities at the Alfv\'en radius along the polar axis are
\begin{itemize}
\item $r_\star = 9.29\,r_o = 0.104$ AU,
\item $V_\star = 103\kmps$.
\end{itemize}
We assume that the observed mass loss rate originates totally from the star. That means the entire mass loss
rate is contained within a flux tube with its last line connected to the star. The last open streamline is adjacent to the 
last closed magnetic field line which defines the magnetosphere.
This assumption yields a mass density, particle density and magnetic field at the Alfv\'en distance along the polar axis of
\begin{itemize}
\item $n_\star = 3.05\times 10^9\cmmt$
\item $\rho_\star = 2.48\times 10^{-15}\gcmmt$
\item $B_\star = 1.82$ G\,.
\end{itemize}
The stellar jet solution can be extended in the disc provided that the disc rotation remains sub-Keplerian. The co-rotation radius, i.e. the radius where the rotation 
frequency of the star equals the Keplerian angular frequency as defined by \cite{Lovelaceetal14}  and \cite{Ferreira13}, is
\begin{itemize}
\item $r_{\rm  corotation} = 1.2 r_\star = 11 \,r_o = 0.13 $ AU
\,.
\end{itemize}
Clearly we are not in the conditions of a propeller. 

The radius of the magnetosphere is slightly smaller than the Alfv\'en radius which in term is  smaller than the corotation radius. In a propeller the radius of the 
magnetosphere is larger than to corotation radius, see \cite{Lovelaceetal14}.

In the simulations presented here, we use a meridional self-similar stellar jet model and extend it to the disc.
We leave  the combination of the stellar jet with a more realistic disc wind for future work.
The stability of the analytical solution is investigated by following the temporal evolution of the standard ideal MHD equations given by the conservation of 
mass, momentum, and energy, together with Faraday's law of induction.
Altogether, they comprise a system of 8 coupled and nonlinear equations that are integrated to obtain
correspondingly the 8 macroscopic MHD quantities, namely the density $\rho$, velocity $\vec V$, magnetic field $\vec B$, and gas pressure $P$,
\begin{equation}
  \frac{\partial\rho}{\partial t} + \nabla \cdot (\rho \vec V) = 0\,,
\end{equation}
\begin{equation}
  \frac{\partial\vec V}{\partial t} + (\vec V \cdot \nabla)\vec V
    + \frac{1}{4\pi\rho}\vec B \times(\nabla\times\vec B)
    + \frac{1}{\rho}\nabla P = - \nabla \Phi\,,
\end{equation}
\begin{equation}\label{eqEnerg}
  \frac{\partial P}{\partial t} + \vec V \cdot \nabla P
    + \Gamma P \nabla \cdot \vec V = H - \Lambda\,=q\,,
\end{equation}
\begin{equation}
  \frac{\partial\vec B}{\partial t} + \nabla \times (\vec B \times \vec V)
    = 0\,.
\end{equation}
The magnetic field satisfies also the divergent-free condition $\nabla\cdot\vec B = 0$.
The gravitational potential is denoted by $\Phi$ and $\Gamma$ is the
polytropic index. Finally, $H$ and $\Lambda$ are the rates of volumetric heating and cooling,
whereas $q$ is the rate of the net volumetric heat flux.

\begin{figure}
  \resizebox{\hsize}{!}{\includegraphics{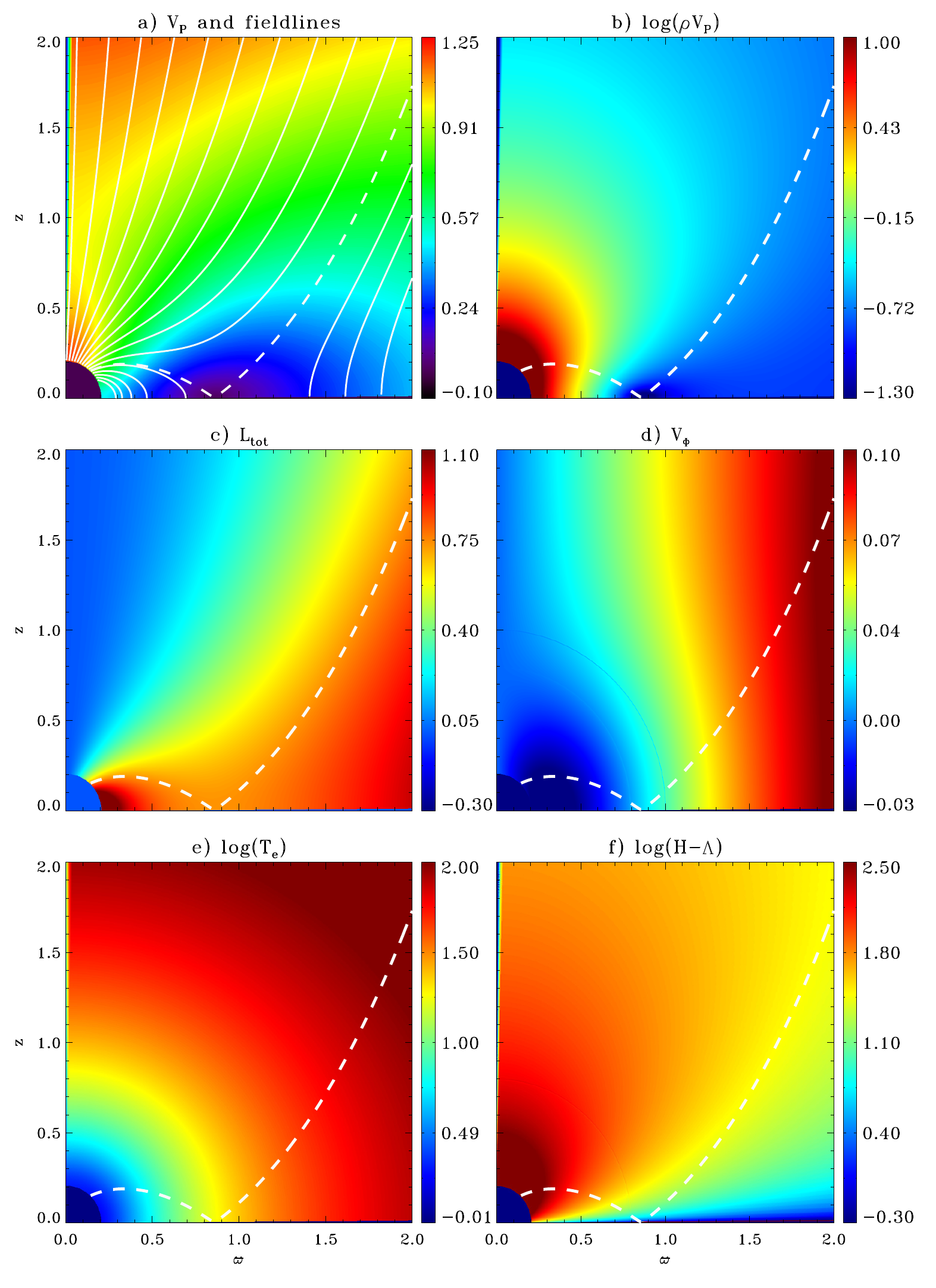}}
  \caption{
    Panel of the stellar jet solution used in \citeauthor{Sautyetal11} (\citeyear{Sautyetal11})
    to describe the RY Tau micro jet.
    In a), we plot the iso-contours of the poloidal velocity together with the streamlines which coincide with the magnetic fieldlines.
    In b) we plot the mass flux density, in c) the
    total angular momentum flux, in d) the toroidal velocity, in e) the effective temperature and 
    in f) the heat flux density. The dashed line corresponds to the last fieldline connected to the star.
      }
\label{Fig1}
\end{figure}

In Fig. \ref{Fig1}, we plot in a panel of six figures, a) the streamlines/fieldlines and the poloidal velocity contours, b) the mass flux, c)
the total angular momentum flux, d) the toroidal velocity distribution, e) the effective temperature, and f) the heat flux density. The dashed line 
is the last open fieldline that is connected to the star. 
  
The magnetic topology of this analytical solution contains three distinct regions. First, a polar region with open fieldlines rooted on the star.
This polar coronal hole is defined by the half opening angle $\theta_{\rm CH}$.
Second, a disc region with open fieldlines rooted on the disc. And third, a region with closed fieldlines, with both footpoints rooted on the
star. The last closed fieldline starts from the inner corona of the star at the point ($r_0$, $\theta_{\rm CH}$) 
and crosses the equator at the X-point ($r_{\rm X}$,  $\pi/2$).
Matter is ejected from the star, as well as from the disc wind part of the outflow model. Our simulations, though, start above the photosphere in the corona 
at $r_0$ which is twice the radius of the star. We are currently performing new simulations starting closer to the star.
These three regions meet at the X-point where the last closed fieldline crosses the equator.
The last open fieldline from the star is the common interface of the three regions.

The toroidal velocity distribution is reversed in the subalfv\'enic equatorial region though the total angular momentum is fixed. For a discussion
on this peculiar phenomenon see \cite{Sautyetal12} and \cite{Cayatteetal14}.
This reversal is a consequence of the strong initial acceleration before the Alfv\'en surface.
However, note that the star is rotating in the $\varphi$ direction and rigidly.

The effective temperature corresponds to the ratio of the pressure over the mass density. The pressure (see Sauty et al., \citeyear{Sautyetal11} for details) may 
include ram or magnetic turbulent pressure, as well as Alfv\'en waves.
It is defined up to a constant and we choose the minimum possible value that gives zero asymptotic pressure and temperature in the jet. 

As mentioned before, the derivation of an analytical, meridional self-similar solution leads to a magnetospheric stationary zone which has non-zero 
poloidal 
velocities parallel to the poloidal magnetic field (see Fig.~\ref{Fig1}). Namely, there is a flow both in the open and the closed magnetic fieldline
regions. The poloidal flow in the closed one is directed towards the
equatorial plane which is thus a sink for the plasma outflow. Obviously, this configuration is not realistic since material cannot pile up 
at the equatorial plane in a steady state. To remedy this problem, we need to suppress the flow in the closed fieldline
region and also change the boundary conditions accordingly. 
We do this numerically, investigating the stability of such modifications.

\subsection{Three cases of boundary and initial conditions in the magnetospheric zone}

The first issue we have to address is how to appropriately modify the boundary and initial conditions of the original closed fieldline region, while 
keeping the basic physics of the analytical solution. We seek to understand how the structure of the magnetospheric zone and the 
streamers may affect the outflow region, independently of the accretion structure. Hence, for the purposes of the present study we will 
assume that, even if a disc is surrounding the star, its accretion rate is negligible. In other words, we assume that all the accreting mass and 
angular momentum is transferred solely to the disc wind. Once we establish the existence of a self-consistent stellar outflow,
with or without a dead zone, we shall proceed to study in a following paper how the accretion affects the central region.  
This is somewhat different from previous studies which concentrate on how accretion affects a simple stellar structure, usually 
close to a dipolar configuration.  

\begin{figure}
%  \resizebox{\hsize}{!}
  \resizebox{.8\linewidth}{!}
  {\includegraphics{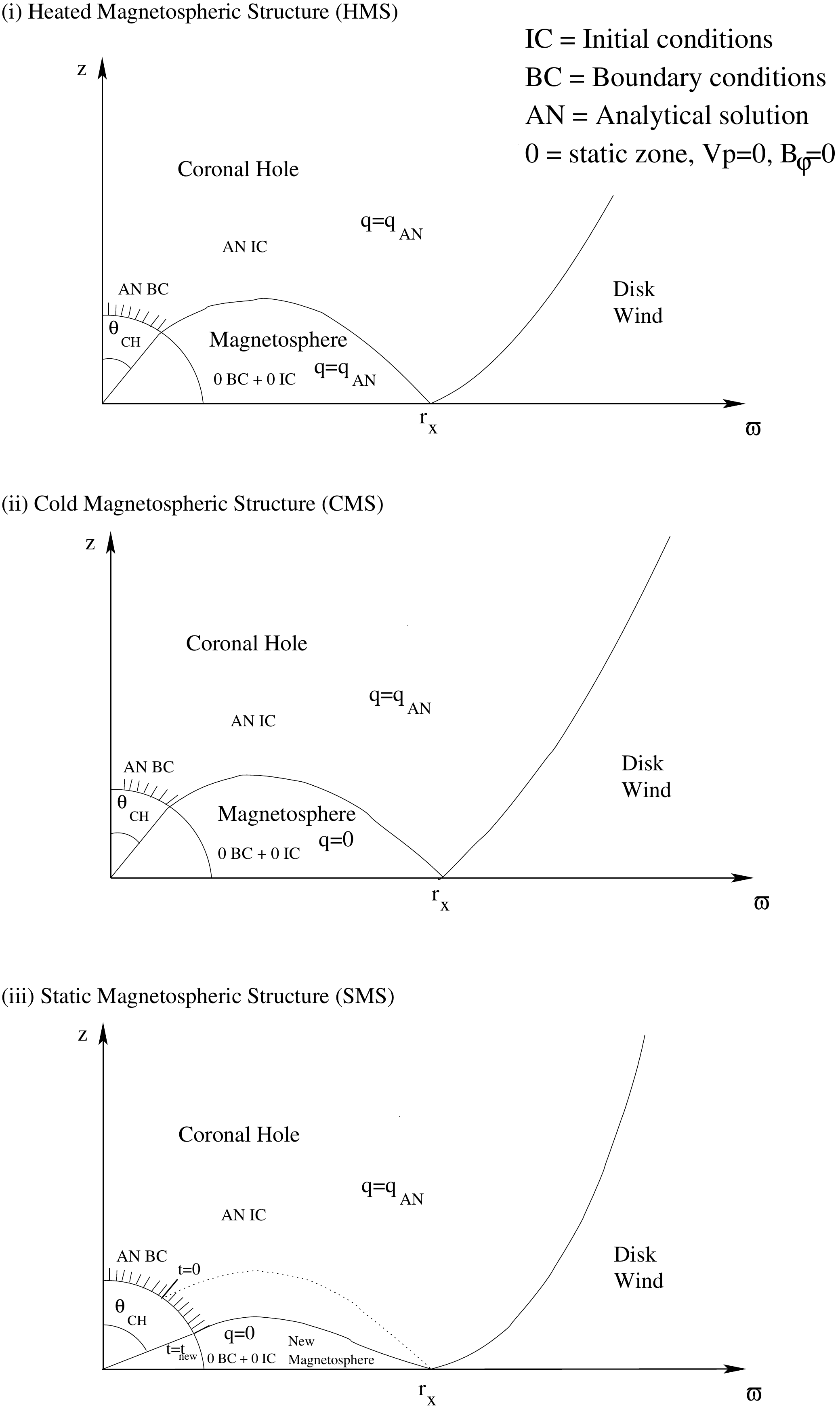}}
  \caption{   
    Plot of the various initial and boundary conditions of the three simulations.
    In (i), the HMS and in (ii), the CMS, the boundary conditions of the polar coronal hole-type region
    are kept fixed to their initial values. In the initially closed fieldline region, the boundary and the initial conditions are set to the ``static equilibrium'' conditions (see text),   
   keeping fixed the heat flux distribution in (i) and a zero heat flux 
    distribution in (ii). In (iii), the SMS, the boundary conditions 
    and the heat flux distribution are those of the analytical solution in the whole open fieldline region.
     The ``static equilibrium'' conditions are applied only in the actual
    closed fieldline region, which evolves with time.
     Note ``0 BC'' stands for ``static boundary conditions'' as explained in the text, while ``0 IC'' stands for ``static initial conditions".
       }\label{Fig2}
\end{figure}

Regarding the initial closed fieldline region, we have chosen to explore three different cases corresponding to three different initial boundary and
initial conditions, which are graphically illustrated in Fig.~\ref{Fig2}.

In the  {\it first} case, the size of the polar coronal hole is fixed between the colatitudes $0$ and $\theta_{\rm CH}$, and corresponds to the initially open 
fieldlines, see
Fig.~\ref{Fig2}(i). 
In this region we keep the boundary conditions of the analytical solution for all physical quantities except density. We let free the  
density but fix its gradient from the analytical solution there.  
In the remaining surface of the star, we assume that the poloidal outflow velocity is zero, or negligible compared to the outflow from the polar coronal
hole region. Thus, we modify the analytical MHD solution imposing $V_p = 0$ in the closed fieldline region. We keep 
this condition along the inner boundary during the simulations, namely, on the stellar surface between the colatitude $\theta_{\rm CH}$ and
$\pi/2$. The equatorial symmetry of the problem ensures that the same condition holds along the equatorial plane between $r_{0}$ and $r_{\rm X}$.

However, imposing $V_p = 0$ in this region has several consequences for other MHD quantities. The projection along $\varphi$ of the momentum  
equation combined with axisymmetry requires that the $\varphi$ component of the Lorentz force should vanish.
This is satisfied if we assume $B_\varphi = 0$ in the closed fieldlines region.
Since the star is considered to rotate almost as a rigid body, we further assume that the closed fieldlines of the magnetosphere are  almost in corotation 
and that $V_\varphi$ varies linearly with distance. The rotation period is around 10 days. This corresponds to a slow rotator with a corotation radius 
beyond the magnetosphere (Lovelace et al., \citeyear{Lovelaceetal14}).
The relations $V_p=0$, $B_\varphi = 0$ and quasi solid rotation constitutes what we shall call the 
``static equilibrium'' conditions because it is static in the corotating frame.

In addition, we keep the heat flux density constant in time in the whole domain of integration and equal to the one of the initial analytical solution.
Thus, Fig.~\ref{Fig2}(i) is referred to as the simulation of a heated magnetospheric structure (HMS).

In the {\it second} case, the so called cold magnetospheric structure (CMS), see Fig.~\ref{Fig2}(ii), all boundary conditions are identical to those of 
Fig. \ref{Fig2}(i), in the polar coronal hole region and along the magnetospheric structure. Additionally, we use the same zero conditions for 
the poloidal velocity and toroidal magnetic field in the initial closed fieldline region. However, the heat flux density is set to zero in this region, such that 
there is no heating.

In the {\it third} case, the so called static magnetospheric structure (SMS), see Fig.~\ref{Fig2}(iii), the boundary conditions and the heat flux distribution
of the analytical solution are kept fixed on all open fieldlines during the simulation, as previously.
The conditions of the closed streamline region are imposed only in the region where the lines are effectively closed.
The difference with the previous cases is that here the angle $\theta_{\rm CH}$ is free to evolve in a self-consistent manner,
and the boundary conditions adapt accordingly.
Thus, the size of the magnetosphere always reduces to its static part.

There is no {\it a priori} reason for these modified field structures to be solutions of the ideal MHD equations. Nevertheless, we assume that those 
configurations can be close to an equilibrium. Our expectation is that the simulation will reach a steady state with a two-component corona, i.e., a 
corona with a wind-zone outside the magnetosphere and a static-zone or a dynamical streamer inside the magnetosphere.
An analogous procedure has been followed in \cite{KeppensGoedbloed99}, who obtained 2.5D solutions for polytropical 
models of a stellar wind with a static zone.
The difference here is that we have a non-polytropic heating function and a different geometry of the flow. Thus, we specify $V_p = 0$, 
$B_{\varphi} = 0$, and a $V_{\varphi}$ that corresponds to a quasi solid rotation.
Regarding the heating, the net heat flux is given by the analytical solution, $H - \Lambda = q_{\rm AN}$. It is applied in the entire
box of the heated magnetospheric case, Fig.~\ref{Fig2}(i). Conversely, we have $H = \Lambda $ in the magnetospheric zone for the second case, Fig.~ 
\ref{Fig2}(ii), and also in the closed fieldline region of the third case, Fig.~\ref{Fig2}(iii).

\section{Numerical modelling}

We carry out the simulations with PLUTO, a numerical code for computational astrophysics (Mignone et al., \citeyear{Mignoneetal07}). All units are
expressed in terms of $r_*$, $v_*$, and $B_*$, which are the Alfv\'enic radius of the analytical solution along the polar axis, and the values of
the speed and the magnetic field there (see Sauty et al., \citeyear{Sautyetal11}). We adopt spherical coordinates and integrate the MHD equations in an
axisymmetric 2.5D domain. 
Our computational box spans in the interval $[r_\mathrm{min},\,r_\mathrm{max}] = [0.2,\,340]$ along the radial direction and
$[\theta_\mathrm{min},\,\theta_\mathrm{max}] = [0,\,\pi/2]$ along the co-latitudinal one. We resolve the radial direction with a total of $512$ cells and in 
particular, a uniform grid from $r = 0.2$ to $r = 2$ ($256$ cells), and a stretched grid for larger radii up to $r = 340$ ($256$ cells).
Along $\theta$, we use $128$ cells which are uniformly spaced. 

For the HMS case, the dead zone has stronger gradients in the magnetic field when compared to the other cases and, 
hence, we double the resolution to 
minimise numerical diffusion. Namely, we use $2048$ cells in the radial direction and $256$ in the latitudinal one.

As initial conditions we set up the analytical stellar wind solution described in \cite{Sautyetal11}.
We modify the region characterised by closed fieldlines according to the descriptions provided in the previous section.
Along the axis we apply axisymmetric boundary conditions, whereas for $r_\mathrm{max}$ we set up outflow conditions.
On the inner boundary ($r_{0}$) we apply the stellar wind conditions for $\theta<\theta_{\rm CH}$, and the static equilibrium conditions 
($V_p = 0$, $B_\varphi = 0$) for larger colatitudes.

For the sake of simplicity, we keep the location of the X point fixed ($r_{\rm X}$) during the simulation.
Physically, this can be justified by assuming that the equatorial equilibrium of the accretion disc
and the stellar magnetic field are not affected by the outflow.
This is a reasonable assumption as long as the stellar magnetic field is stronger than the magnetic field of the disc.

The simulations are carried out up to a final time when the axisymmetric MHD wind has converged to a state which
is considered to be stationary.
This typically happens within a few crossing times of the distance from the stellar surface to the Alfv\'en point,
a feature that has been also observed in many previous numerical studies of self-similar solutions.

\section{Physical interpretation of the results of the simulations}
\label{Sec.magnetospheric_wind}

\begin{figure}
  \resizebox{\hsize}{!}{\includegraphics{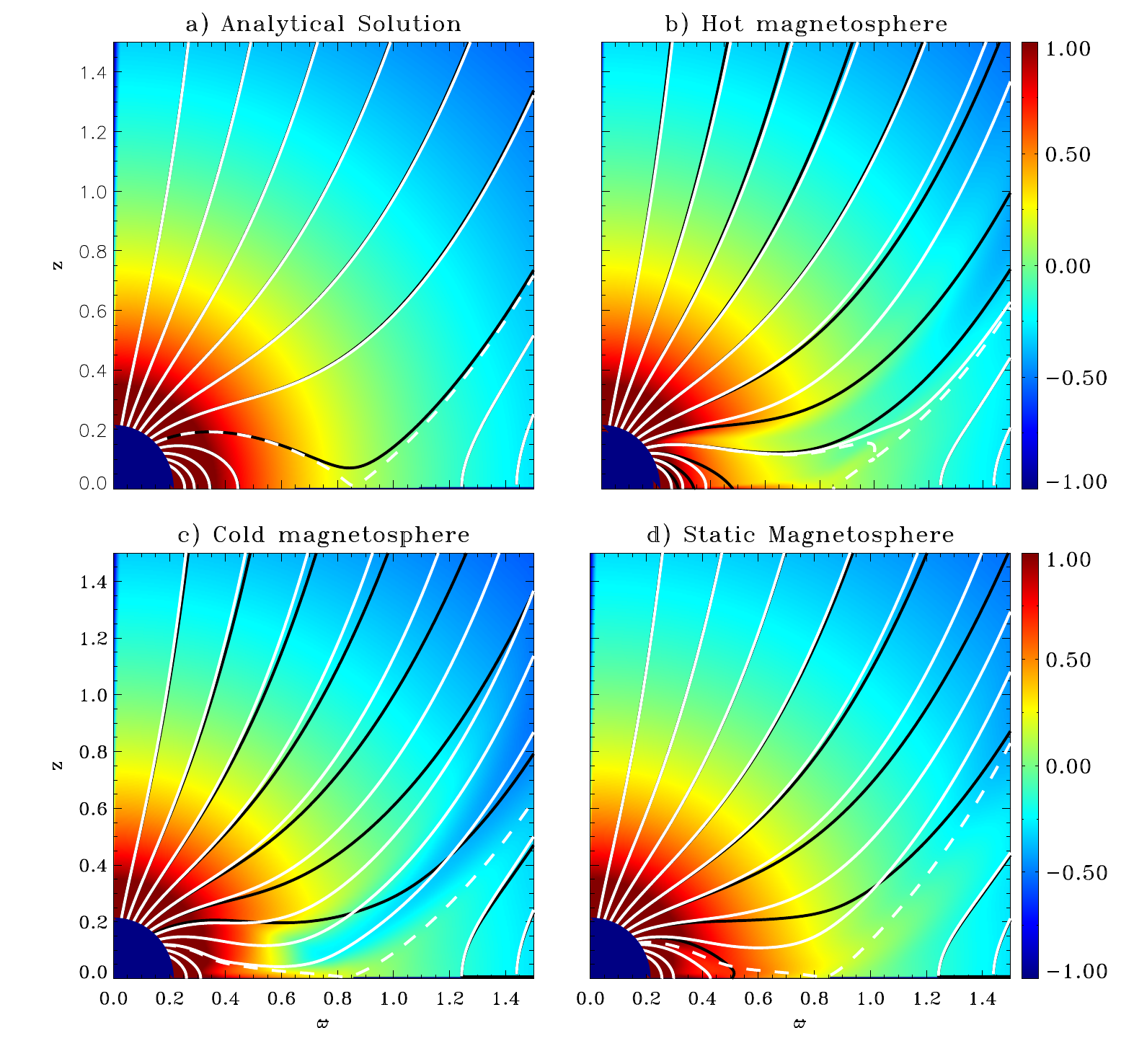}}
  \caption{
  We plot the density contours in logarithmic scale of the final steady state of each of the four cases,
  the analytical solution (a), the HMS (b), the CMS (c) and the SMS (d). 
  The magnetic fieldlines are plotted with white solid lines and the flow streamlines with black.
  The white dashed line corresponds to the last open stellar fieldline.
   } \label{Fig3}
\end{figure}

In Figs. \ref{Fig3}, \ref{Fig4} and \ref{Fig5}, we plot the density, total angular momentum and poloidal velocity distributions
for the following cases,
\begin{enumerate}
\item
the initial analytical solution, 
\item
the heated magnetosphere (HMS), 
\item
the cold magnetosphere (CMS) and finally 
\item
the static magnetosphere (SMS). 
\end{enumerate}
In particular, Fig.~\ref{Fig3} shows the density contours, the magnetic fieldlines (white solid lines), and the streamlines (black solid lines).
Fig.~\ref{Fig4} displays the total angular momentum flux and the magnetic fieldlines, whereas,
in Fig.~\ref{Fig5} we plot the poloidal velocity isocontours and the streamlines.

Finally, Fig.~\ref{Fig6} summarises the fieldlines (white solid lines) and streamlines (dark solid lines) together 
with the boundary conditions used in each case. The yellow sector on the star indicates regions where the stellar boundaries are given by the analytical 
solution, while the green sector corresponds to the magnetospheric conditions. The same applies along the equatorial plane. 
In dark blue we have shaded the region with no net heat flux and in light blue the region where we have kept the net heat flux from the analytical 
solution.   The red thick line indicates the last opened streamline/fieldline in the beginning of the simulation ($t=0$). The orange lines 
delimit the streamer  region when it exists (Figs. \ref{Fig6}b and c), and the purple one, the region where the fieldlines are closed 
(Figs. \ref{Fig6}b, c and d).

From these plots, we may deduce the stability of the stellar outflow independently of the formation
of a pure static dead zone or a slow outflowing streamer or both.
In the following, we compare and discuss the different cases in detail.

\subsection{Topological stability of the polar stellar outflow and stationarity}

In all numerical experiments we have performed, the stellar jet around the polar axis is remarkably stable and maintains its initial steady state. 

By keeping the boundary conditions of the analytical solution on the coronal hole, as well as the analytical heat flux density within the
computational domain, the polar stellar outflow is found to be topologically stable.
In addition, all MHD quantities keep their initial distributions globally. 
This may be seen by comparing the density and poloidal speed of Fig.~\ref{Fig1} with those of Figs.~\ref{Fig3} and \ref{Fig5}, respectively.
The final configurations are practically identical to the initial one in the coronal hole.
For the SMS case, panel (d) of each figure, this implies that the unaffected region is larger, since the coronal hole is slightly bigger.

We also note the relative steadiness of the polar outflow. This can be deduced for the parallelism of the vector fields and the conservation of angular 
momentum. Around the polar axis, the poloidal magnetic fieldlines are parallel to the poloidal streamlines, with only small deviations as we move away 
from it.
In addition, the total angular momentum isocontours are almost parallel to the poloidal magnetic fieldlines
of the coronal hole (see Fig.~\ref{Fig4}). 

This proves the topological stability of the initial analytical solution despite the non steady magnetospheric region.
Moreover, notice that the streamers observed in the HMS and the CMS cases, as well as the outflowing plasma at the border of the disc wind
in the SMS case, are time dependent and the transported angular momentum variable.
However, this does not affect the robustness of the polar coronal hole ejection. 

\subsection{Formation of a coronal streamer between the polar stellar outflow and the disc-wind region in the hot magnetosphere simulation (HMS)}
\label{Sec.HMS}

In the HMS case, the initial heating of the analytical solution remained fixed throughout the simulation.
Note that this heating induces poloidal flows along the magnetic fieldlines
of the magnetospheric zone, that converge towards the equator.

\begin{figure}
  \resizebox{\hsize}{!}{\includegraphics{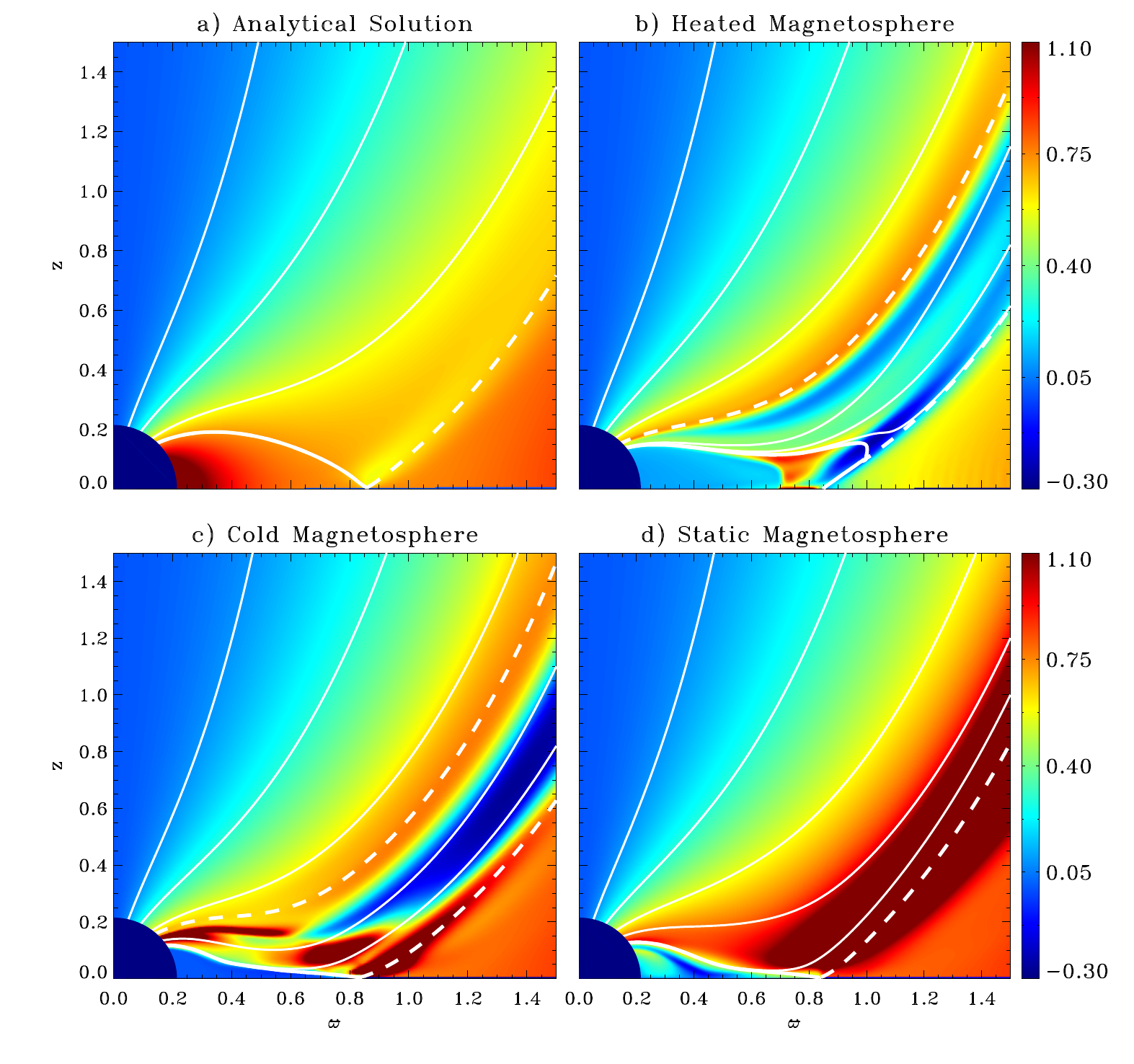}}
  \caption{
  Panel of the final angular momentum flux for the four cases as in Fig.~\ref{Fig3}. 
  The analytical solution in a), the HMS in b), 
  the CMS in c) and the SMS in d).
  The white solid lines correspond to the magnetic fieldlines. In b) and c) the upper dashed line is the magnetic boundary of the coronal hole, 
  while the bottom
  one delimits the disc wind. In a) and d) those two dashed lines  are coincident and cross the X-point.
   } \label{Fig4}
\end{figure}

\begin{figure}
  \resizebox{\hsize}{!}{\includegraphics{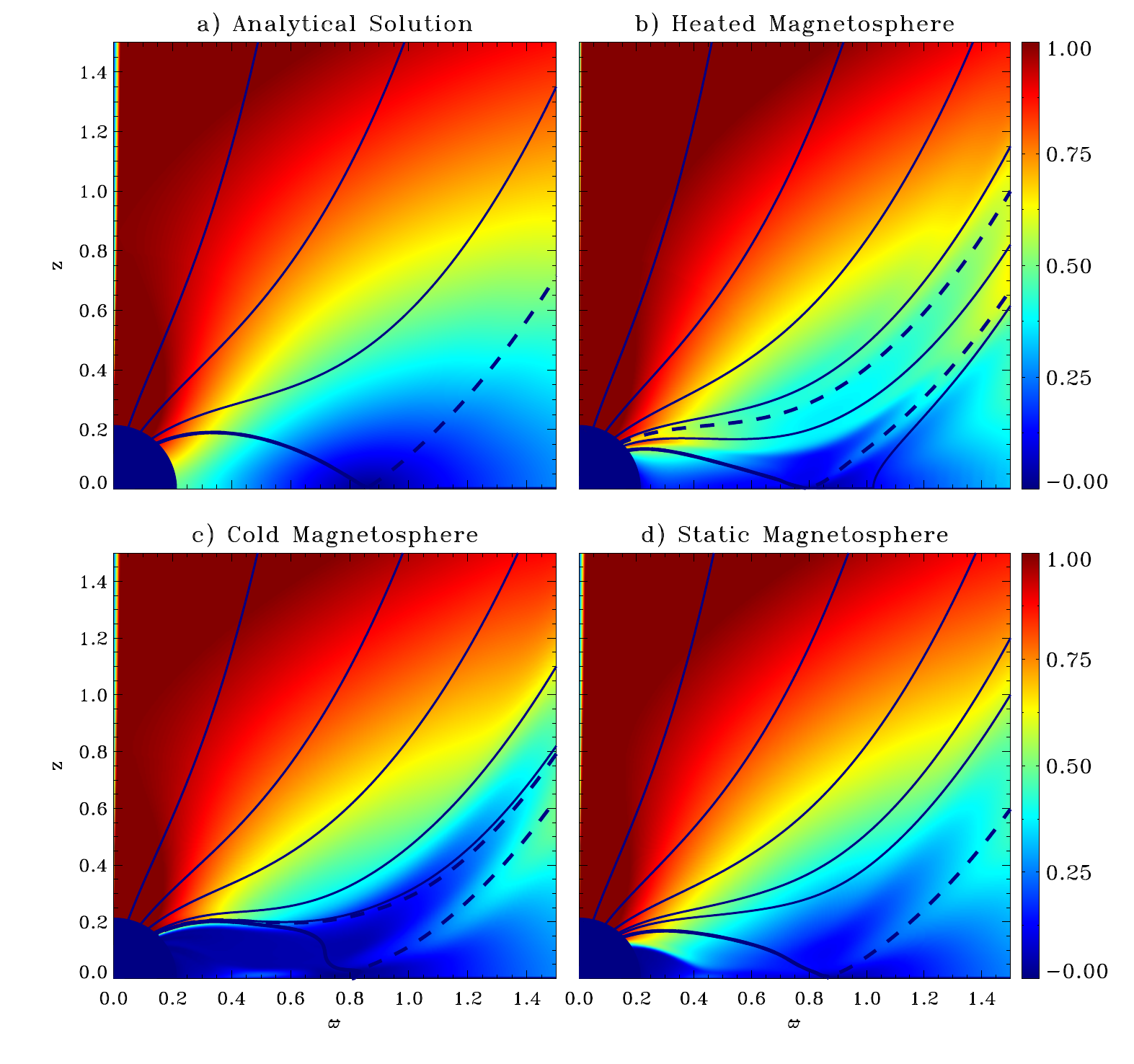}}
  \caption{
  Contours of the poloidal velocity magnitude for the four simulations placed as in Fig.~\ref{Fig3}. 
  The analytical solution in a), the HMS in b), the CMS in c), and the SMS in d).
  The dark solid lines are the streamlines of the flow which give the direction of the poloidal velocity.
  In b) and c) the upper dashed line is the flow boundary of the coronal hole, while the bottom
  one delimits the disc wind. In a) and d) those two dashed lines  are coincident and cross the X-point.
   } \label{Fig5}
\end{figure}

Since we have switched off this poloidal flow along the closed magnetic fieldlines, the heated atmosphere in the initially
closed magnetic fieldlines region expands (see the final state of the HMS simulation in Figs.~\ref{Fig3}b, \ref{Fig4}b and \ref{Fig5}b).
However, this region is bounded by the disc wind and, as a result, the last closed magnetic fieldline is pushed outwards tangentially to the disc wind.
This feature is best seen in Fig. \ref{Fig4}b where the deformed magnetosphere bulges towards an upper right direction.
In turn, this leads to the formation of a so called streamer region between the two orange lines of Fig. \ref{Fig6}b or, equivalently, the two white dashed 
lines seen in Fig. \ref{Fig4}b. 

As seen in Figs. \ref{Fig3}b, \ref{Fig5}b and \ref{Fig6}b, the streamlines of the inner part of the streamer are almost perpendicular to the closed 
magnetic fieldlines.
This implies that the configuration is dynamical and not steady. The flow carries away magnetic flux from the 
star, which is additional evidence that we are away from a steady state. 
Thus, a few of the closed magnetic lines of the initial magnetosphere open and we obtain a 
streamer-type configuration. The streamer is limited by the two dashed lines of Fig. \ref{Fig4}b, between the coronal ejection and the disc wind. On the 
same plot, we show the last closed fieldline, which bulges before opening at a ``Y'' point of reconnection along the first line rooted onto the disc. 

In this region where these fieldlines are squeezed (between the two dashed fieldlines of Fig. \ref{Fig4}b), the total angular momentum flux $L$ becomes 
lower due to the fact that we impose on the stellar boundary $B_{\varphi}=0$.  
Hence, the average angular momentum flux of the streamer is lower  than the one of the surrounding areas of the stellar outflow and  disc wind (dark blue 
colour).
The dark red triangle just below the reconnection point is a consequence of the magnetic flux accumulation before the lines open. 
Furthermore, combining Fig. \ref{Fig3}b and Fig. \ref{Fig5}b, we see that the mass flux is lower between the two dashed fieldlines because $V_p$
is set to zero at the stellar boundary of the initial magnetosphere. 

Altogether, in the HMS case we obtain,
\begin{enumerate}
\item
a coronal hole type configuration around the polar axis with a fast outflow, 
\item
the disc wind region and 
\item
a streamer type configuration with a slower flow at larger polar angles.
\end{enumerate}
The flow lines of the streamer are squeezed between the last open fieldline of the initial magnetosphere and the separatrix between the initial stellar
wind 
region and the disc wind region.  The streamer type configuration is a flow of low density and velocity which carries away angular momentum and 
magnetic flux in a non steady way. It creates an area favourable to the formation of an X or "fan" wind (Shu et al., \citeyear{Shuetal94}; 
Ferreira \& Casse, \citeyear{FerreiraCasse13}) provided that mass flux comes from the accretion disc. 


\subsection{Formation of a dead zone and a low density streamer between the polar stellar outflow and the disc-wind in the cold magnetosphere 
simulation 
(CMS)}
\label{Sec.CMS}

\begin{figure}
\resizebox{\hsize}{!}{\includegraphics{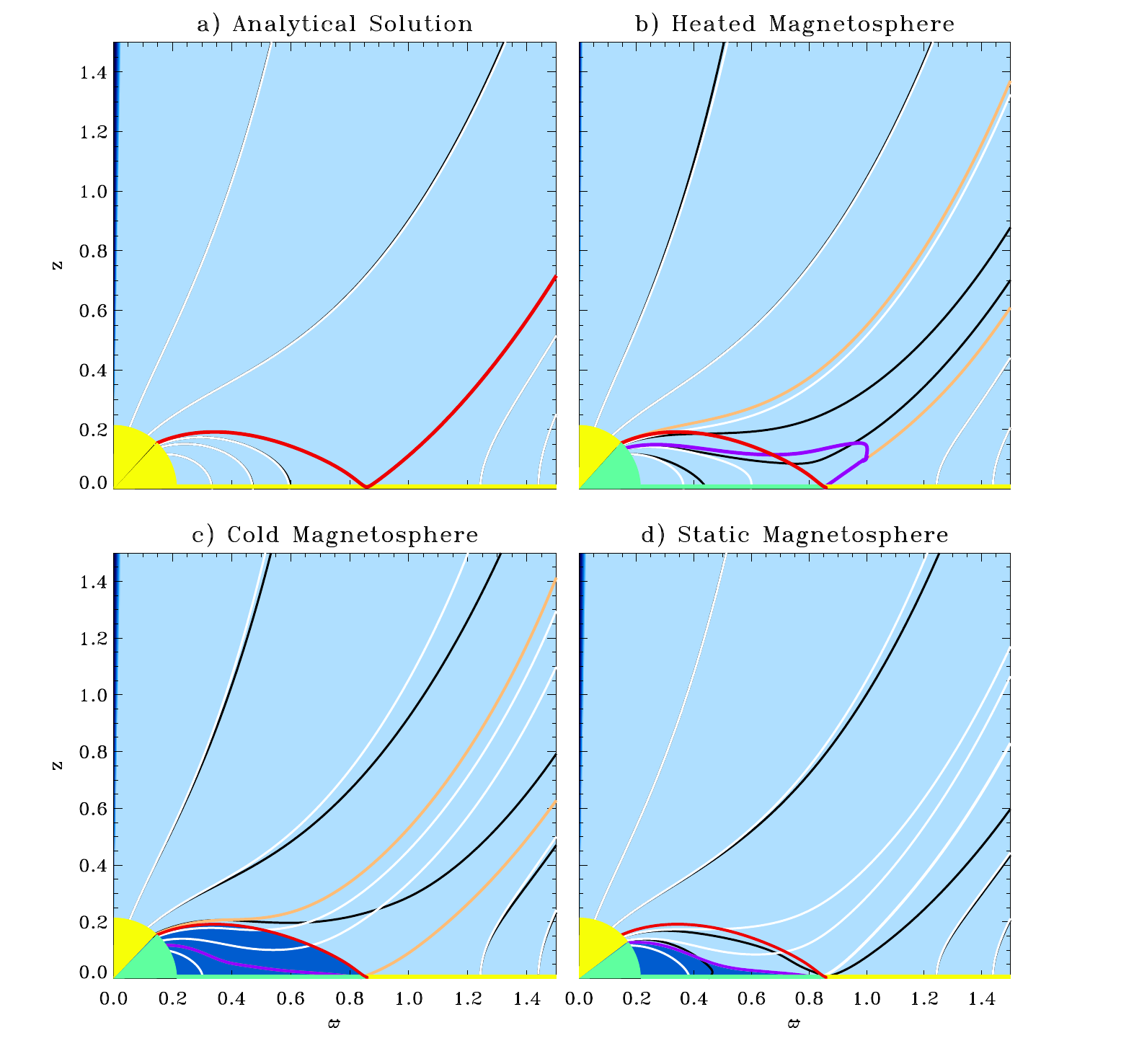}}
\caption{
  Selected streamlines (dark solid lines) and magnetic fieldlines (white solid lines) for
  the final configuration of the simulations, also highlighting the boundary conditions of the four cases.
  The panels are placed as in Fig.~\ref{Fig3}, the analytical solution in a), the HMS in b), the CMS in c), and the SMS in d). 
  The yellow sectors on the star and the yellow lines along the equatorial plane indicate the location 
  where the coronal hole boundary conditions are imposed. The green ones correspond
  to the magnetospheric conditions of ''static equilibrium'', $V_{\rm p}=0$ and $B_\varphi=0$. The light blue shade corresponds to the region where
  the net heat flux is given by the analytical solution, while the dark blue corresponds to a zero net heat flux.
  The red thick line indicates the last stellar stream/fieldline of the initial conditions.
  The orange lines delimits the streamer region and the purple one the dead zone region (see text). 
    }\label{Fig6}
\end{figure}

The CMS case differs from the previous HMS case only in the magnetosphere, which here has a zero net heat flux (the dark blue region seen in 
Fig. \ref{Fig6}c, compared to Fig. \ref{Fig6}b). Despite the different heating assumed, we note that a streamer region of a similar size and 
morphology forms here too. Again, the streamer region is the flow between the two orange lines of Fig. \ref{Fig6}b or, equivalently, the two white dashed 
lines seen in Fig. \ref{Fig4}b. 

In the streamer region, the plasma pushes against the fieldlines and flows out, but
with a smaller velocity and mass flux due to the absence of heating (Figs.~\ref{Fig3}c and \ref{Fig5}c).

However, the CMS case shows an interesting new feature, the presence of a static zone in the magnetosphere that can be defined as a dead 
zone. The dead zone is characterised by closed magnetic fieldlines, an almost zero velocity (see Fig.~\ref{Fig5}c) and streamlines that are directed 
towards the inner part of the magnetic fieldline region  (see Figs.~\ref{Fig3}c and \ref{Fig6}c, and compare \ref{Fig4}c with \ref{Fig5}c). Thus, even 
though the simulation has not reached a perfect steady state, the dead zone can be considered as a stationary feature. 

Note that the actual dead zone is considerably smaller in size than the initial closed magnetosphere. The possible physical reasons could be the 
following.  {\it First}, by setting the poloidal velocity equal to zero there is no inertial force towards the exterior of the magnetosphere. {\it Second}, by 
setting $B_{\varphi} =0$, the magnetic pressure is reduced inside the magnetosphere  and {\it third}, by switching off the heating inside the 
magnetosphere, there is a deficit of the internal gas pressure. As a result the inner total pressure is lower than the external one. 

Such a dead zone was not obtained with the previous case of the HMS in which there was a heating ($q \neq 0$), even though the other two imposed 
conditions were the same, namely $B_{\varphi}=0$ and $V_p=0$. Thus, it is the absence of heating that prevents the inner pressure to increase and 
allows for a partial static  equilibrium. In other words, by switching off the heating inside the initial closed magnetosphere, the closed fieldline region 
shrinks until it counterbalances the external pressure. On the contrary, in the HMS case the imposed heating builds up an unceasing pressure which 
blows up the magnetic field. 
On top of that, the numerical reconnection contributes to the opening of the magnetic fieldlines. 

At  this point it is remarkable that both simulations (HMS and CMS) have a comparable streamer structure of similar size and which coincides with the 
opening of the initially last closed fieldlines (Fig.~\ref{Fig6}). This suggests that the size and the morphology of the streamer depend only on the 
boundary conditions which are identical for both simulations, and not on the heat flux. To further understand this point we calculate the Alfv\'en Mach 
number and the plasma beta of both cases. In the region of the streamers, both quantities remain much smaller than unity throughout, implying that the 
geometry is dominated by the magnetic field and is not affected by the flow dynamics or by the pressure. 
The only exception is around the ``Y'' point of the HMS case, where both the Alfv\'en Mach number and the plasma beta are greater than 
unity. Nevertheless, this is expected because of the null point of the magnetic field where the reconnection occurs. 

The streamer of the CMS is dynamically different from that of the HMS. The absence of 
heating limits the pressure driving of the flow. As a result, the net mass flux (and density) in the CMS case is slightly higher close to 
the star -- and much lower further out -- than in the HMS case (Figs.~\ref{Fig3}b and ~\ref{Fig3}c).
Consistently, the velocity of the CMS streamer is everywhere lower than in the HMS (Figs.~\ref{Fig5}b and ~\ref{Fig5}c).
As a consequence, the angular momentum of the CMS case accumulates close to the star and is not transported along the streamer
(Fig.~\ref{Fig3}c) contrary to the HMS case (Fig.~\ref{Fig3}b).

Altogether, in the CMS case we obtain,  
\begin{enumerate}
\item
a coronal hole type configuration around the polar axis with a fast outflow,
\item
the disc wind region,   
\item
a low density streamer type configuration with a slower wind at larger polar angles,
and with its streamlines squeezed between the last open fieldline 
of the initial magnetosphere and the last closed fieldline of the dead zone,
\item
a smaller closed fieldline region than the initial one but which is a static dead zone.
\end{enumerate}

\subsection{Formation of a larger coronal polar stellar outflow and a dead zone in the static magnetosphere simulation (SMS)}
\label{Sec.SMS}

The final distributions of the density,  specific angular momentum, poloidal velocity, and global topology of the SMS case are shown in 
Figs.~\ref{Fig3}d, \ref{Fig4}d,  \ref{Fig5}d  and \ref{Fig6}d.

As in the previous two cases (HMS, CMS), the initial configuration of  these quantities is maintained in the polar portion of the outflow. 
However, as the closed fieldline region shrinks during the simulation, we readjust the boundary and heating conditions in the open fieldline one.
Namely, in the streamer zone we replace the so called "static equilibrium" conditions (zero poloidal velocity, zero toroidal magnetic field and quasi solid 
rotation at the stellar boundary) and the no heat flux in the domain with the so called "polar coronal hole conditions" (the boundaries are given by the 
analytical solution, and the heat flux is included within the flux tubes of open fieldlines). 

Here, the formation of a streamer region, as defined in the previous cases of CMS and HMS, is suppressed.
However, we maintain the conditions in the dead zone and the system reaches a quasi static equilibrium.
The flow in the coronal region at the end of the iterative process is remarkably close to the analytical solution, 
except in a layer close to the dead zone.
In this region, the flow is not completely steady but the velocity is rather small and the dynamics evolve very slowly.

Thus, this simulation demonstrates that an appropriate modification of the analytical solution can reach an equilibrium with a polar coronal fast wind and 
an equatorial dead zone.
The outer fieldlines of the initial magnetosphere open up and matter accelerates along them due to the heating that has been reintroduced in the 
simulation. As a consequence, the vicinity of the last connected line is emptied from the accumulated material.
The specific angular momentum also increases along the former closed fieldline region because of the heating we have put back ($q \neq 0$) and the 
release of the zero azimuthal field condition ($B_{\varphi} \neq 0$ there, see Fig.~\ref{Fig5}d).

Despite the lack of a streamer similar to the CMS case, note that the dead zone is remarkably similar in size and morphology to the dead zone obtained in 
the CMS case. Again, this is due to the magnetic field dominating the transfield equation, regardless of the dynamics of the surrounding outflow.
In particular, apart from the vicinity of the X equatorial point, the plasma beta and the Alfv\'en Mach number are smaller than unity in this case too.
This confirms that the dead zone is topologically stable for a given set of boundary conditions. 

The shrinking of the dead zone can be explained in a similar manner to the CMS case.
Since $B_\varphi$ has been set to zero initially, the dead zone has a magnetic
pressure deficiency. In addition, the absence of net heat flux in the closed fieldline region also creates a lack of gas pressure. 
As a result, an overall inwards force due to the total pressure will be exerted on its boundary that will
tend to squeeze the dead zone in time.
Furthermore, the current sheet that is generated by the discontinuity of the
toroidal component of the magnetic field at the interface, as well as the shear that develops due to the velocity gradients,
trigger reconnection events.
In turn, the outer fieldlines of the dead zone open up, launching the
enclosed plasma. Depending on whether the dead zone is being heated or not, that region will either
consist of a pressure driven outflow, or remain vacuum, respectively.

The final state can be divided in {\it four} regions, 
according to whether or not the initial conditions are maintained. 
\begin{enumerate}
\item
The {\it first} region is the stellar outflow zone from the polar coronal hole. Globally, the stellar wind keeps its initial topology and is not affected in
any significant way by the change in the conditions in the initial magnetospheric zone.
\item
The {\it second} region is the border of the coronal zone. The analytical boundary conditions and heating are maintained, i.e., heating $q =q_{\rm AN}\neq 0$, 
$B_{\varphi} \neq 0$ and $V_p\neq 0$, which open up the initially closed fieldlines and accelerate matter along them like in the polar 
part. However, this accelerated wind is less steady than the polar one.
The fieldlines and the streamlines are not yet parallel (Fig. \ref{Fig6}d) and the angular momentum is not constant
(Fig. \ref{Fig4}d). Nevertheless, the variations are much smaller than in the streamer regions of the two previous cases. 
This configuration is reminiscent of an X-wind or at least can favour its formation. Future simulations that include
accretion will lead to a more clear characterization of this part of the outflow and also examine whether the properties
of the so-called X-wind appear in such outflow configurations.

\item
The {\it third} part of the configuration is the dead zone, which lies within the initial magnetospheric zone.
Its fieldlines have a deformed topology but similar to the initially closed fieldlines.
One of the most important features of this case is the fact that the zone with closed fieldlines has a smaller extent
than in the initial setup.
\item
The {\it fourth} region of the field is the disc wind.
Its deviation from the initial configuration is very small and the structure stays qualitatively the same. 
\end{enumerate}

\section{Discussion}

\subsection{The distribution of mass and magnetic fluxes}

\begin{table}
\tiny
\begin{tabular}{llcc}
% \resizebox{.8\linewidth}{!}
%\renewcommand{\arraystretch}{2.5}
\multicolumn{4}{c}{\bf Mass Fluxes} \\
\hline
Solution type   & location                & Mass Stellar Flux & Outer Mass flux along \cr
and zone      & of the starting flow                   & from the Star & the same magnetic lines  \cr
                  &                                  & (in units of $2\pi\rho_\star V_\star r_\star^2$)& (in units of $2\pi\rho_\star V_\star r_\star^2$)  \cr
\hline
Analytical solution (Figs. a), zone 1 & $R=0,2$ , $0<\theta<\theta_{\rm CH}$                                       & 0.50             & 0.50    \cr
Analytical solution (Figs. a), zone 2 & $R=0,2$ , $\theta_{\rm DZ}<\theta<\pi/2$                                   & 0.60             & 0.59    \cr
\hline 
Heated magnetosphere (Figs. b), zone 1 & $R=0,2$ , $0<\theta<\theta_{\rm CH}$                               & 0.50             & 0.45   \cr
Heated magnetosphere (Figs. b), zone 2 & $R=0,2$ , $\theta_{\rm DZ}<\theta<\pi/2$                          & 0.12            & 0.10\cr
Heated magnetosphere (Figs. b), zone 3 & $R=0,2$ , $\theta_{\rm CH}<\theta<\theta_{\rm DZ}$        & 0.05             & 0.13\cr
\hline
Cold magnetosphere (Figs. c), zone 1 & $R=0,2$ , $0<\theta<\theta_{\rm CH}$                                  & 0.48                   & 0.42 \cr
Cold magnetosphere (Figs. c), zone 2 & $R=0,2$ , $\theta_{\rm DZ}<\theta<\pi/2$                              & -0.01  $\pm$ 0.01         & 0.04\cr
Cold magnetosphere (Figs. c), zone 3 & $R=0,2$ , $\theta_{\rm CH}<\theta<\theta_{\rm DZ}$             & 0.02         & 0.07\cr
\hline
Static magnetosphere (Figs. d), zone 1 & $R=0,2$ , $0<\theta<\theta_{\rm CH}$                                  & 0.66             & 0.52 \cr
Static magnetosphere (Figs. d), zone 2 & $R=0,2$ , $\theta_{\rm DZ}<\theta<\pi/2$                              & 0.02             & 0.12\cr
\end{tabular}
\caption{We calculate the mass fluxes in three given fieldline flux tubes in two domains each. 
The inner domain corresponds to the mass flux injected from the star 
in the flux tube, the outer domain is the mass flux exiting from the same flux tube. The three zones are defined by the surface of the star where the 
magnetic fieldlines of the flux tube are rooted. Zone 1 coincides with the coronal hole stellar surface. zone 2 corresponds to the closed magnetic fieldlines, 
which connect the star to the equatorial plane. Finally, zone 3 is the streamer region, i.e. the small region of open magnetic fieldlines where we have 
static boundaries. Zone 3 exists only in the two cases of the heated and cold magnetospheres. 
}\label{tabmassflux}
\end{table}

\begin{table}
\tiny
\begin{tabular}{llcc}
\multicolumn{4}{c}{\bf Magnetic Fluxes}\\
\hline
Solution type   & location                & Magnetic Stellar Flux & Outer magnetic flux along \cr
and zone      & of the starting flow                   & from the Star & the same magnetic lines  \cr
                  &                                  & (in units of $B_\star r_\star^2$)& (in units of $B_\star r_\star^2$)  \cr
\hline
Analytical solution (Figs. a), zone 1 & $R=0,2$ , $0<\theta<\theta_{\rm CH}$                                       & 0.49           & 0.49      \cr
Analytical solution (Figs. a), zone 2 & $R=0,2$ , $\theta_{\rm DZ}<\theta<\pi/2$                                   & 0.57             & 0.56    \cr
\hline 
Heated magnetosphere (Figs. b), zone 1 & $R=0,2$ , $0<\theta<\theta_{\rm CH}$                               & 0.49            & 0.46    \cr
Heated magnetosphere (Figs. b), zone 2 & $R=0,2$ , $\theta_{\rm DZ}<\theta<\pi/2$                          & 0.49            & 0.59   \cr
Heated magnetosphere (Figs. b), zone 3 & $R=0,2$ , $\theta_{\rm CH}<\theta<\theta_{\rm DZ}$        & 0.11            & 0.22   \cr
\hline
Cold magnetosphere (Figs. c), zone 1 & $R=0,2$ , $0<\theta<\theta_{\rm CH}$                                  & 0.47             & 0.44       \cr
Cold magnetosphere (Figs. c), zone 2 & $R=0,2$ , $\theta_{\rm DZ}<\theta<\pi/2$                              & 0.47           & 0.41       \cr
Cold magnetosphere (Figs. c), zone 3 & $R=0,2$ , $\theta_{\rm CH}<\theta<\theta_{\rm DZ}$             & 0.13         & 0.27        \cr
\hline
Static magnetosphere (Figs. d), zone 1 & $R=0,2$ , $0<\theta<\theta_{\rm CH}$                                  & 0.66            & 0.60  \cr
Static magnetosphere (Figs. d), zone 2 & $R=0,2$ , $\theta_{\rm DZ}<\theta<\pi/2$                              & 0.38           & 0.43  \cr
\end{tabular}
\caption{We calculate the magnetic fluxes that come in and out from the same three given fieldline flux tubes as in Tab. \ref{tabmassflux}.  Zone 1 is the 
stellar corneal hole region, zone 2 is the closed fieldline region and zone 3 coincides with the streamer. 
}\label{tabmagflux}
\end{table}

We calculate the mass and magnetic fluxes in the following three regions of the final configurations of each simulation. First,
the stellar outflow from the coronal hole (zone 1), second the closed magnetosphere (zone 2), and third the streamer (zone 3), when it exists.
Our purpose is to compare the new equilibria with the analytical solution and evaluate how much the final steady states
deviate from that solution.
In addition, by calculating the fluxes at two different locations, one at the stellar surface and one further away,
we can estimate the amount of material that crosses the interfaces which define the three aforementioned regions.
This allows us to check how far we are from steadiness as this would imply that the mass flux should be conserved downstream.
The reference value for the mass flux is $3\cdot10^{-9}M_\odot /$yr, which is the one of the analytical solution
in agreement with the value of \cite{Sautyetal11}.
In the following, the mass flux is given in units of $2\pi\rho_\star V_\star r_\star^2$ and the magnetic flux is given in units of $B_\star r_\star^2$.
We give the results with two significant digits but this is already at the limit of precision of our analysis and far below what can be constrained by 
observations.

First, we calculate the mass flux across the open fieldlines of the coronal hole (zone 1) for the four simulations.
We integrate the flux over two surfaces, one at the stellar surface, where the open streamlines are anchored,
and one at the upper part of the plotted domain, i.e. the horizontal edge of the box plus the upper part of the right vertical edge.

For the closed field region (zone 2), we use the same procedure.
The inner surface is the spherical cap at the base of the flow where  the closed fieldlines are rooted. The outer surface
is the region along the equatorial plane from the star up to the X point.
The second region coincides with the dead zone in the CMS and SMS cases (c and d), and to the closed fieldlines that have a flow or streamer for the 
Analytical and HMS cases (a and b).

For the magnetospheric streamer (zone 3), when it exists, the inner surface is the region with open fieldlines but also with "static equilibrium''
boundaries, which means the same fieldlines were closed at the initial of the simulation. The outer surface is the vertical one at the right edge of the box
for the same lines.

We do not calculate the fluxes from the disc wind as we know that the disc wind solution should eventually be replaced by a more realistic one.

As it can be seen from Tables~\ref{tabmassflux} and \ref{tabmagflux}, both the mass and the magnetic fluxes
of the coronal hole zone of the Analytical, HMS and CMS cases (a, b, and c respectively) are very close to the analytical solution already published.
In the last case only (Fig.~\ref{Fig6}d), the corresponding fluxes are higher because the coronal hole is $\sim$20\% larger. Thus, the fluxes are also
around 20 to 30\% higher.
Notice also that the mass flux of the SMS (d) is lower at the outer surface when compared to the inner surface of integration. This
implies that mass is crossing the interface separating zones 1 and 2.
It shows up in the bottom right panel of Fig.~\ref{Fig3}. We obtain the same results for the HMS (b) and the CMS (c) even if this is less obvious. 
In all three cases this is because the edge of the coronal hole is not completely in a steady state. The flow lines of the coronal
hole edge cross the fieldlines and transfer mass to the outer region.

Regarding zone 2, the magnetosphere of the HMS (b) has a non zero mass flux due to the heating that accelerates the flow. This just reflects that the 
closed fieldline region of this case is not in a quasi static equilibrium but is in reality very dynamical.
Simulations for the CMS and the SMS (c and d respectively) have an almost zero value for the mass flux, consistently with the formation of the static dead 
zone. A significant increase of the mass flux appears in the SMS case (d) in the outer domain of integration of zone 2. 
Mass flux from the coronal hole still mixes with the mass flux of the dead zone because we have not reached perfect steady equilibrium.

For all three cases (b, c, d), the outer surface of zone 2 has a non zero value although it is small. This is just a signature of the weak remaining flow that 
converges towards the equator, because we calculate the fluxes on the line of cells above the equatorial boundary. 
This flow is sliding along the equatorial plane and feeds the 
light plumes merging from the X-point, seen in yellow on Figs.~\ref{Fig3} (excess of mass density) and in light blue on Figs.~\ref{Fig5} (excess of flow 
speed).
The magnetic flux is close to the analytical solution, as expected. There is  a small excess of the outer magnetic fluxes in the HMS (b) and the SMS (d) 
cases because of the loss of magnetic flux through reconnection. The CMS (b) case has a deficit of magnetic flux probably because the magnetic
reconnection occurs in the streamer region only.

Finally, the streamer (zone 3), when it exists (Fig. \ref{Fig6}b and Fig. \ref{Fig6}c), appears to have higher fluxes at the outer integration surface.
This suggests that mass and magnetic field from the coronal hole  and the closed fieldline region are crossing the interfaces that separate these two zones
from  the streamer. 
This occurs due to the fact that the simulations for the HMS and CMS (b and c respectively) are not exactly in a steady state, but they are still slowly 
evolving. This also means the streamer is taking away magnetic flux via reconnection at the X-point. 

We did not recalculate the angular momentum loss. However, considering the small variation of the stellar mass loss and magnetic fluxes,
the average braking time we estimated in our previous paper should not change drastically.
Namely, these self-consistent numerical models suggest that the stellar wind can brake the star in approximately one million years.

\subsection{Summary and conclusions}

The main motivation for the work reported in this paper, has been to produce a self-consistent MHD model with an outflow along open polar magnetic fieldlines, 
surrounded by an equatorial dead zone of closed magnetic field lines. 
To achieve this goal, we extended an analytical MHD stellar wind model with a polar outflow and an equatorial magnetosphere.
Starting from the initial analytical solution, we modified the equatorial magnetospheric zone by imposing realistic initial conditions,
i.e., zero poloidal velocity and zero toroidal magnetic field.
We performed {\it four} different numerical simulations to investigate the stability of the stellar outflow. Our aim has been to find 
a self-consistent global MHD solution that includes an equatorial dead zone along with the polar wind.

The {\it first} simulation we have carried out corresponds to the original analytical solution (case a).
In the {\it second} simulation, we switched all initial magnetospheric flows to zero, but kept the heating distribution of the analytical solution (case b, Hot Magnetosphere or 
HMS). The {\it third} case (c), Cold Magnetosphere or CMS, is similar to case (b) but the heating within the region of the initial magnetosphere is switched off.
{\it Finally}, case (d) (Static Magnetosphere or SMS), is the most physically relevant. It builds on top of simulation (c) by adapting the boundary conditions and the 
heating to the evolution of the magnetosphere.

Our conclusions can be summarized as follows : 
\begin{enumerate}
\item
The stellar outflow emerging from the polar coronal hole regions is maintained in all cases and remains close to the original analytical solution.
\item
When the heating in the magnetosphere is maintained as in the original solution, but  the poloidal flow is switched off, 
an increase of the thermal pressure results inside the magnetosphere 
from Bernoulli's law.
In turn, its outer magnetic fieldlines open up and form a streamer along which material accelerates and escapes.
This thin low density region is sandwitched between the polar stellar wind and the disc wind (case b, HMS).
\item
When the heating inside the magnetosphere is turned off, a  squeezed dead zone forms, due to the deficit of internal total pressure.
The region of the magnetic fieldlines that open up does not have the required energy to drive any outflowing plasma and hence the streamer
that forms is almost empty (case c, CMS).
\item
In case (d), SMS, the appropriate readjustments of the boundary conditions and of the heating, consistently with the size of the dead zone,
lead to the expansion of the plasma in the coronal hole regions.
In particular, the outer magnetic fieldlines of the magnetosphere open up and effectively enlarge the base of the stellar outflow.
An equatorial dead zone forms and is found to coexist in a steady state with the polar stellar outflow.

The formation of the above last configuration suggests that the extension of the analytical stellar wind MHD model to incorporate a consistent dead zone is 
possible.
As a result, the obtained numerical solution maintains  the physical properties of the original analytical one, such as, the mass
and angular momentum loss rates.
\end{enumerate}

\section*{Acknowledgments}

This work has been supported by the project ``Jets in young stellar objects:
what do simulations tell us?'' funded under a 2012 \'{E}gide/France-FCT/Portugal
bilateral cooperation Pessoa Programme and by the Scientific Council of Paris
Observatory (PTV program).
We acknowledge financial support from "Programme National de Physique Stellaire" (PNPS) of CNRS/INSU, France,
from Fundacao para a Ci\^encia e a Tecnologia (FCT), Portugal (research grant UID/FIS/04434/2013)
and from FEDER through COMPETE2020 (POCI-01-0145-FEDER-007672).
KT acknowledges the hospitality and support of the Laboratoire Univers et
Th\'eories, Observatoire de Paris, during his visit supported by the Scientific
Council of Paris Observatory (PTV program).
TM was supported in part by NASA ATP grant NNX13AH56G.

\label{lastpage}
 
\end{document}